# Object permanence in newborn chicks is robust against opposing evidence

Justin N. Wood[1-3*], Tomer D. Ullman[4-5], Brian W. Wood[1], Elizabeth S. Spelke[4-5], & Samantha M. W. Wood[1,3*]

[1]Informatics Department, Indiana University Bloomington; Bloomington, IN.
[2]Cognitive Science Program, Indiana University Bloomington; Bloomington, IN.
[3]Neuroscience Department, Indiana University Bloomington; Bloomington, IN.
[4]Department of Psychology, Harvard University; Cambridge, MA.
[5]Center for Brains, Minds, & Machines, MIT; Cambridge, MA.

*Corresponding authors. Email: woodjn@indiana.edu or sw113@indiana.edu

Newborn animals have advanced perceptual skills at birth, but the nature of this initial knowledge is unknown. Is initial knowledge flexible, continuously adapting to the statistics of experience? Or can initial knowledge be rigid and robust to change, even in the face of opposing evidence? We address this question through controlled-rearing experiments on newborn chicks. First, we reared chicks in an impoverished virtual world, where objects never occluded one another, and found that chicks still succeed on object permanence tasks. Second, we reared chicks in a virtual world in which objects teleported from one location to another while out of view: an unnatural event that violates the continuity of object motion. Despite seeing thousands of these violations of object permanence, and not a single non-violation, the chicks behaved as if object permanence were true, exhibiting the same behavior as chicks reared with natural object permanence events. We conclude that object permanence develops prenatally and is robust to change from opposing evidence.

**One-sentence summary:** We raised newborn chicks in worlds where object permanence was either true or false, and found that chicks always behave as if object permanence is true.

People expect objects to persist and move continuously across space and time. This cross-species expectation of "object permanence" is regarded as a core foundation for learning (*1–6*). However, the origins and nature of object permanence remain heavily debated. The debate can be broken into two specific questions. First, when does object permanence develop: Is this expectation present at birth, or is it the result of postnatal development, learned from experience as newborns see objects moving in and out of view? Second, what is the nature of object permanence: Is it flexible, continuously adapting to the statistics of experience, or is it robust to change, even in the face of opposing evidence?

Prior studies of object permanence have entirely focused on the first question. Studies of object permanence began with Piaget (*7*), who concluded that object permanence does not develop until 8-9 months of age, based on highly replicable findings that younger infants fail to search for hidden objects. More recently, however, researchers have found evidence for object permanence in younger infants, using simplified reaching tasks (*8–11*) and object tracking tasks focused on looking time (*12–15*). These studies indicate that infants have some expectation of object permanence within the first few months of life, much younger than Piaget theorized.

The strongest evidence that object permanence develops early in life comes from controlled-rearing studies on newborn chicks. Unlike newborn humans, newborn chicks can move around their environment, allowing them to search for hidden objects. There is evidence for object permanence in chicks as early as 4-5 days post-hatching (*16–20*). For example, 5-day-old chicks that were trained to search for their imprinted object behind a screen were able to find the object when it disappeared behind one of two screens (*16*). Similarly, when 5-day-old chicks were imprinted to a virtual object, they searched for the object in the correct location after it disappeared behind one of two virtual screens (*18*). These studies suggest that object permanence can be present and functional at birth, during an animal's first encounters with objects.

But what is the nature of this initial object knowledge (2nd question)? While both infant and controlled-rearing studies demonstrate that object permanence is present early, these findings are compatible with two different explanations for its nature. On one account, object permanence is flexible, continuously adapting to the statistics of experience. Studies across psychology (*21-23*) and neuroscience (*24-25*) have shown that visual systems learn through statistical learning, associating features that co-occur in the input stream. Human infants and adults, for example, extract statistical regularities from sensory input to construct higher-order object concepts (*26-28*). There is evidence that visual systems can change rapidly (in a matter of hours) when presented with altered visual statistics (*29*), suggesting that object knowledge might be flexible.

On a second account, object permanence is rigid and robust to change. Prior to birth, brains generate spontaneous waves of retinal activity, and these object-like patterns could predispose newborn animals to perceive the world in terms of enduring objects that persist over space and time (*30, 31*). If these retinal waves occurred during critical periods, which are widespread during early brain development (*32-35*), then it is possible that early expectations of object permanence could become enduring and resistant to change.

Both accounts (flexible vs. rigid) make the same predictions when animals are reared in natural worlds. Natural worlds provide continuous evidence that objects move on connected spatiotemporal paths, so both a flexible and robust system would be expected to have object permanence. The flexible system should have object permanence because it adapted to the visual statistics in the world (*37-39*). The rigid system should have object permanence because it was grown and trained during a critical early period, during which initial knowledge was "frozen into place."

To distinguish between these accounts, we must systematically alter the experiences available to newborn animals, then measure whether those experiences alter the animal's object permanence behavior. To this end, we performed VR-based controlled-rearing studies on newborn chicks (Fig. 1A).

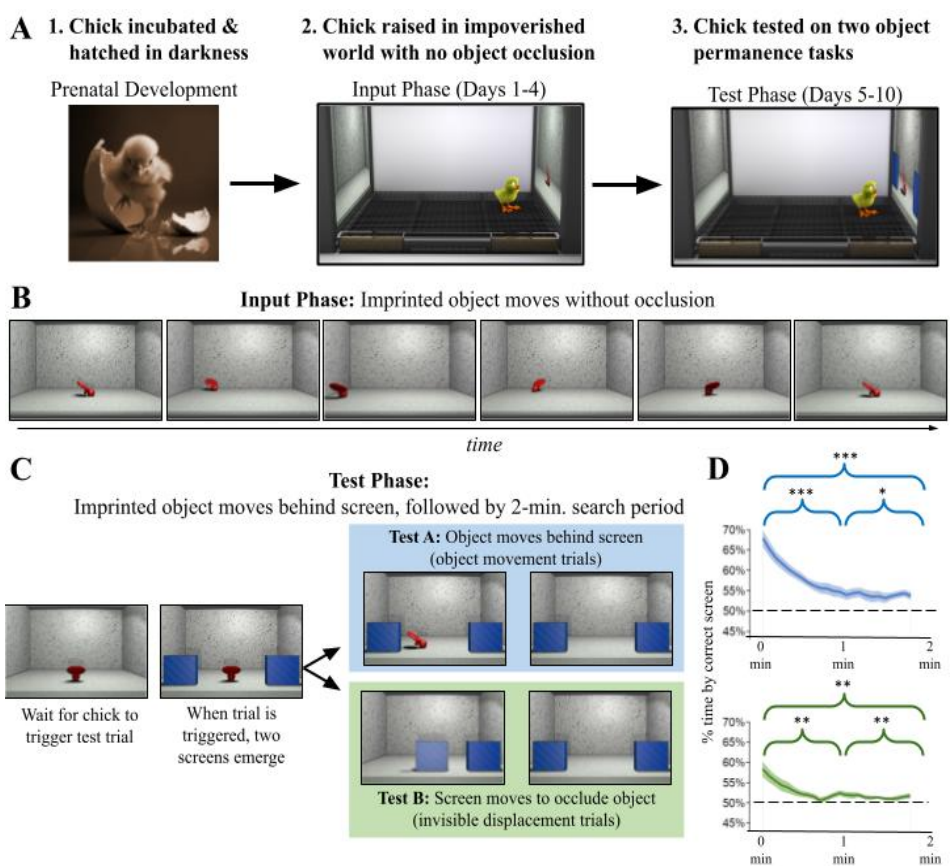

**Fig. 1.** Experiment 1. **(A)** Newborn chicks were hatched in darkness, raised in impoverished worlds with no object occlusion events, then tested on object permanence tasks. **(B)** In the Input Phase, the chambers contained a single object moving in a virtual room. The environment contained no occluders, so objects never moved in and out of view from behind other objects. **(C)** In the Test Phase, we used a chick-directed design to measure the chicks' object permanence expectations across two tasks: object movement (*blue*) and invisible displacement (*green*). **(D)** In both tasks, the chicks spent significantly more time near the correct versus incorrect screen during the 2-min. search period after the object moved out of view. Significance levels (*$p$ ≤ .05, **$p$ ≤ .01, ***$p$ ≤ .001).

**Experiment 1: Development of object permanence in 'impoverished' worlds**

Before testing whether object permanence is flexible vs. robust to opposing evidence, it was first necessary to create an experimental system that allowed newborn animals to be raised continuously (24/7) in virtual worlds. To do so, we combined the automated controlled-rearing method previously used to study newborn chicks (*18*) with a video game engine (Unity 3D). The game engine controlled the virtual environment surrounding the chick, allowing them to trigger object hiding events when they moved next to the virtual object. We used a game-based testing platform because it is automated (preventing experimenter error and bias), interactive, and flexible (*36*).

In Exp. 1, we verified that this new game-based controlled-rearing method could produce interpretable findings, consistent with prior studies that used real objects and manual testing methods. These prior studies showed that newborn chicks have object permanence expectations on their first encounters with objects (*16-20*). Thus, we tested whether our method could reproduce this finding in a virtual world, with fully automated testing procedures.

We tested whether newborn chicks develop object permanence when they are reared in impoverished visual worlds that contain no object occlusion events. In the Input Phase (Days 1-4), each chick was reared in a chamber that contained one virtual object moving around a virtual room (Fig. 1B, SI Movie 1). The chamber contained no other objects (virtual or real), so the object never moved in and out of view from behind another object. Based on prior work showing that chicks imprint to virtual objects (*40–42*), we expected the chicks to imprint to the virtual object, then attempt to reunite with the object when separated.

In the Test Phase (Days 5-10), we used a forced-choice procedure to test the chicks' object permanence (Fig. 1C). To ensure that test trials were only triggered when the chick was active and interested in their imprinted object, we used a chick-directed design. At the beginning of each test trial, the imprinted object appeared on the opposite side of the chamber from the chick. When the chick moved to reunite with the object, they triggered a test trial (via automated image-based tracking). When the trial was triggered, two virtual screens emerged from the floor of the virtual room, one on each side of the object. Then, one of two hiding events occurred: an "object movement trial" or an "invisible displacement trial."

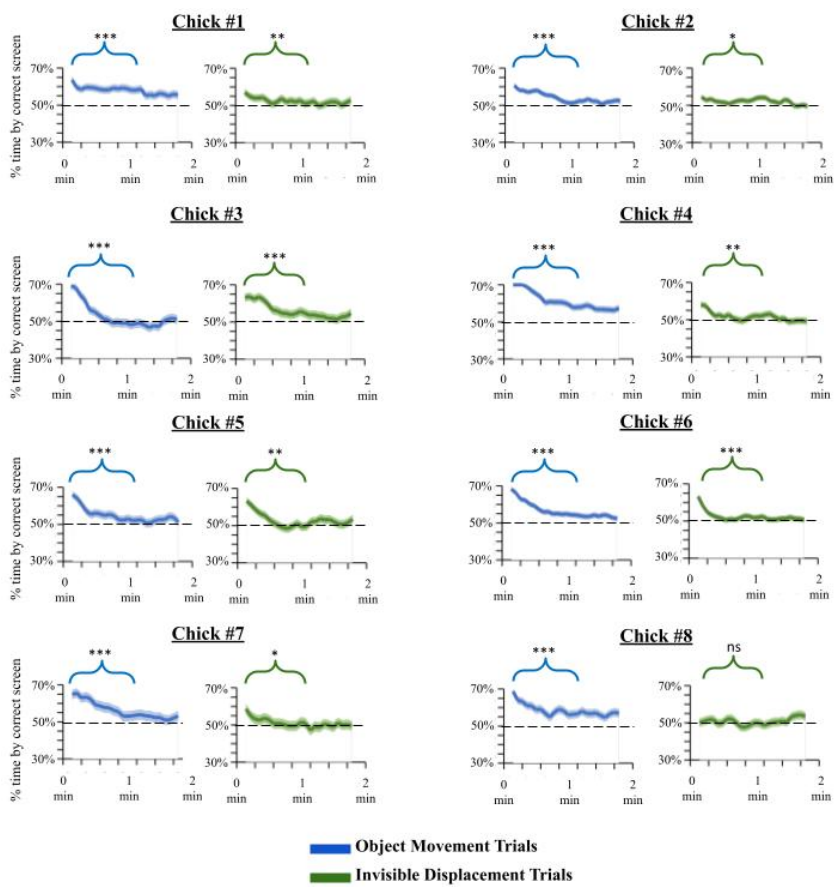

**Fig. 2.** Individual subject performance in Exp. 1. On the object movement trials (*blue*), all of the chicks (8/8 subjects) spent significantly more time near the correct versus incorrect screen during the first minute of the search period. On the invisible displacement trials (*green*), most of the chicks (7/8 subjects) spent significantly more time near the correct versus incorrect screen during the first minute of the search period. Significance levels: *$p ≤ .05$, **$p ≤ .01$, ***$p ≤ .001$.

On the object movement trials, the imprinted object turned toward one of the screens, then moved behind the screen until it was fully occluded (SI Movie 2). On the invisible displacement trials, one of the screens moved to the center of the room to occlude the imprinted object. Then, the screen (with the object still behind it) moved back to its original position (SI Movie 3). So, on the invisible displacement trials, the chicks needed to infer the position of their imprinted object based on the movement of the screen. Invisible displacement tasks are more difficult than object movement tasks, since the hidden object must be tracked as it moves from its initial hiding position to its final hiding position (*7*).

In both events, the object remained hidden for two minutes. Once the object moved behind the screen (object movement trials) or the screen returned to its original location (invisible displacement trials), we measured whether the chick spent more time near the screen hiding the imprinted object.

The chicks demonstrated an object permanence expectation across the entire search period, spending significantly more time near the screen hiding the object. This pattern was observed on both the object movement trials (one-sample *t*-test, $t(7) = 6.52$, $p = .0003$, Cohen's $d = 2.31$) and the invisible displacement trials (one-sample *t*-test, $t(7) = 4.10$, $p = .005$, Cohen's $d = 1.45$).

To provide a more fine-grained analysis of behavior, we analyzed performance as a function of the amount of time the object was hidden from view (Fig. 1D). We found that performance was high at the beginning of the search period, then gradually declined as the object remained hidden for more time. In the first minute, the chicks spent more time by the correct screen than the incorrect screen, on both the object movement trials (one-sample *t*-test, $t(7) = 8.29$, $p = .0001$, Cohen's $d = 2.93$) and the invisible displacement trials (one-sample *t*-test, $t(7) = 3.87$, $p = .006$, Cohen's $d = 1.37$). The chicks showed a weak preference for the correct screen during the second minute of the search period, indicating that the chicks could remember the location of the object for at least one minute.

We also examined whether the overall pattern of success on object permanence tasks held at the individual chick level, by leveraging the >700 test trials collected per chick (Fig. 2). We found that on the object movement trials, all chicks spent significantly more time near the correct screen in the first minute of the search period (one-sample *t*-test for each chick, all *P*s < .00003). On the invisible displacement trials, all chicks but one spent more time near the correct screen in the first minute of the search period (one-sample *t*-test for each chick, all *P*s < .05).

Exp. 1 shows that newborn chicks expect objects to continue to exist after moving behind other objects, despite receiving no visual evidence that this is the case. This finding replicates prior studies showing that newborn chicks have object permanence expectations during their first encounters with objects. This experiment also shows that our game-based testing method can produce robust and interpretable results.

**Experiment 2: Development of object permanence in 'unnatural' worlds**

In Exp. 2, we tested whether object permanence is flexible versus robust against opposing evidence, by creating virtual worlds that either followed natural or unnatural rules (Fig. 3A). In the Input Phase (Days 1-4), we reared chicks with one virtual object that continuously moved in and out of view from behind two screens (Fig. 3B). In the natural world, when an object moved behind a screen, it always reemerged from behind that same screen, consistent with object permanence (SI Movie 4). In the unnatural world, when an object moved behind one of two screens, it always reemerged from behind the opposite screen, inconsistent with any continuous path of motion (SI Movie 5). Both events were equally predictable: in the natural world, the object always reemerged from behind the same screen; in the unnatural world, the object always reemerged from behind the opposite screen.

The chicks were presented with thousands of these natural or unnatural object occlusion events across the Input Phase. If object permanence is flexible, then only the chicks reared in the natural world should develop object permanence. But, if the expectation of object permanence is robust to change, including persistent violations of spatiotemporal continuity, then newborn chicks should have object permanence expectations in both the natural and unnatural worlds.

In the Test Phase (Days 5-8), the chicks saw similar hiding events as in the Input Phase, except that the screens never lowered to reveal the location of the object (Fig. 3C, SI Movies 6-7). Thus, the test trials provided no information about whether objects move continuously when hidden from view. Once the object had moved out of view, we measured the amount of time the chicks spent next to each of the two screens for 2 min. We coded the "natural screen" as the screen that the object moved behind (i.e., the screen where the object would be in a natural world), and the "unnatural screen" as the screen the object did not move behind (i.e., the screen where the object was hidden in the unnatural world). If the chicks raised in the unnatural world learned to predict the outcome of the unnatural hiding events, then they should have spent more time near the unnatural screen than the natural screen on the test trials.

We found that chicks spent significantly more time by the natural screen, regardless of whether they were reared in the natural or unnatural world (one-sample *t*-tests; natural world: $t(3) = 4.58$, $p = .02$, Cohen's $d = 2.29$; unnatural world: $t(3) = 4.01$, $p = .03$, Cohen's $d = 2.00$).

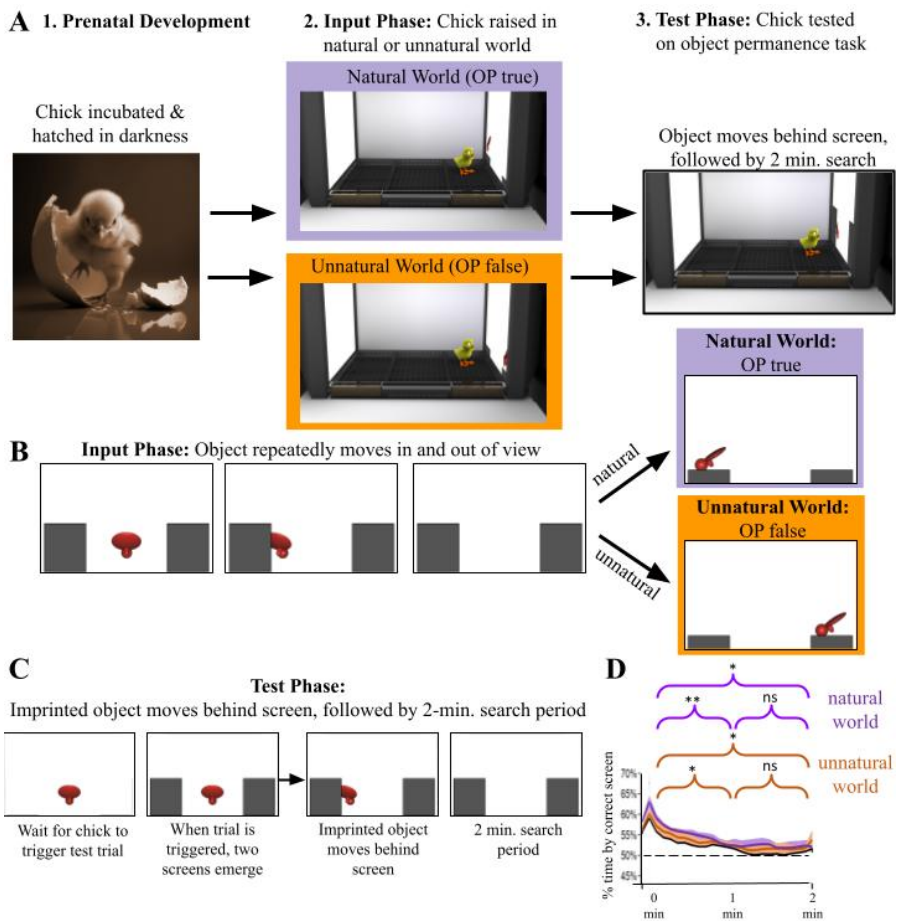

**Fig 3.** Experiment 2. **(A)** Newborn chicks were hatched in darkness, raised in either natural (*purple*) or unnatural (*orange*) worlds (where object permanence was true or false, respectively), then tested on an object permanence task. **(B)** In the Input Phase, the chambers contained a single object continuously moving in and out of view. In the natural world, the object always reappeared from behind the same screen; in the unnatural world, the object always reappeared from behind the opposite screen. **(C)** In the Test Phase, we used a chick-directed design to measure the chicks' object permanence expectations. **(D)** The chicks spent significantly more time near the natural versus unnatural screen during the search period, regardless of whether they were reared in the natural versus unnatural world (significance levels: *p ≤ .05, **p ≤ .01, ***p ≤ .001).

As in Exp. 1, performance was high at the beginning of the search period and then gradually declined (Fig. 3D). In the first minute, the chicks spent more time by the natural screen, both when raised in the natural world ($t(3) = 11.18$, $p = .002$, Cohen's $d = 5.59$) and the unnatural world ($t(3) = 4.97$, $p = .02$, Cohen's $d = 2.48$).

As in Exp. 1, we also examined individual chick performance (Fig. 4). We found that on the first minute of the search period, all the chicks spent more time by the natural versus unnatural screen, regardless of whether they were raised in the natural world (one-sample $t$-tests, all $P$s < $10^{-5}$) or unnatural world (one-sample $t$-tests, all $P$s < .01).

Overall, the chicks behaved as if objects persist over occlusion and move continuously, independent of whether they had received visual evidence that was consistent or inconsistent with this behavior. Object permanence in newborn chicks appears robust to change from opposing evidence.

**Discussion**

We used VR-based controlled rearing to explore the origins and nature of object permanence. In Exp. 1, we reared newborn chicks in a virtual world that contained no object occlusion events and found that chicks still solve object permanence tasks. This study replicated prior studies reporting that newborn chicks have object permanence expectations (*16-20*). In Exp. 2, we reared chicks in an unnatural virtual world in which object occlusion events occurred repeatedly, but when an object was occluded, it always reappeared in a location that could not be reached by any continuous motion. Despite seeing thousands of these violations of object permanence, and not a single non-violation, the chicks behaved as if object permanence were true. In fact, the chicks reared in the unnatural world developed nearly identical behavior to the chicks reared in a natural virtual world. Object permanence is both present early in life and robust to change from opposing evidence.

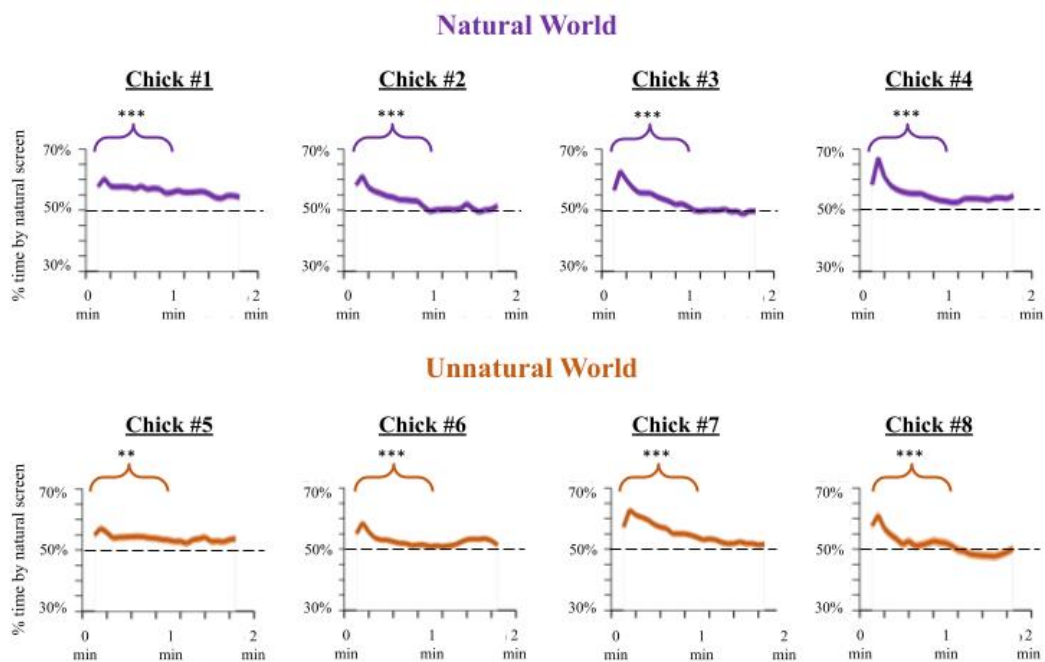

**Fig. 4.** Individual subject performance in Exp. 2. All of the chicks (8/8 subjects) spent significantly more time near the natural versus unnatural screen during the first minute of the search period, regardless of whether they were reared in the natural world (*purple*) versus unnatural world (*orange*). Significance levels: *$p$ ≤ .05, **$p$ ≤ .01, ***$p$ ≤ .001.

Our findings provide evidence that object permanence is a core inductive bias for visual perception. Traditionally, two signatures characterize an inductive bias: 1) it allows systems to make inferences that go beyond the training data, and 2) it constrains the range of input-output functions that can be learned (*43–46*). Our experiments provide evidence for both signatures in newborn chicks. First, chicks succeed on object permanence tasks prior to acquiring object permanence experience (Exp. 1). The outputs of object perception go beyond the training data received from the environment. Second, chicks fail to learn about unnatural events where hidden objects move on unconnected spatiotemporal paths (Exp. 2). So, there are constraints on the types of visual events that can be learned.

Additional evidence for a core object-based inductive bias comes from controlled-rearing studies of object recognition. During their first encounters with objects, newborn chicks can parse objects from backgrounds (*47*), bind colors and shapes into integrated object representations (*48*), and build invariant object representations that generalize across new viewing situations (e.g., changes in view, background, motion speed, and motion direction (*49-52*). Thus, soon after hatching, newborn chicks build robust, abstract, and enduring object representations.

How might prenatal processes build an object-based inductive bias? At least two processes predispose newborn brains toward perceiving the world in terms of spatially and temporally continuous objects. First, prenatal brains generate spontaneous waves of activity in the retina and other parts of the cortex long before eye opening (*53–56*). These waves are bounded, cohesive units that move on connected spatiotemporal paths (*57*). As such, the earliest training data available to developing visual systems are inherently "object-like."

Second, brains employ a learning rule that favors temporal continuity (*58*): spike timing-dependent plasticity (STDP). This learning rule potentiates synapses based on the temporal correlation between spikes in pre- and postsynaptic neurons. Simulations show that when these two processes (retinal waves and STDP) are combined in self-organizing neural networks, the networks spontaneously develop priors for solving visual perception tasks (*59, 60*). Since retinal waves and STDP are present during prenatal development in both birds and mammals (*61–63*), an object-based inductive bias may serve as a cognitive prior for newborns across the animal kingdom.

Developing brains also move through sensitive and critical periods. These periods slow or halt learning, preventing some brain structures from continuing to change as a function of experience. We speculate that critical periods cause some of the robustness against opposing evidence that we observed in chicks.

**Which experiences shape learning?** We are not claiming that object permanence is robust against all types of experience. It is an open question which experiences matter, and which do not, for developing, maintaining, and enriching object knowledge. For instance, controlled-rearing studies show that experience with *visible* objects moving on slow, continuous paths is essential for the development of object recognition (*64-66*) and object permanence (*18*). When chicks are reared with objects that move on visibly non-smooth (discontinuous) paths, the chicks develop impaired object permanence. Thus, visible violations of spatiotemporal continuity 'break' object permanence expectations, whereas the hidden violations of spatiotemporal continuity tested here had no impact on chicks' object permanence behavior.

Ultimately, VR will be an essential tool for characterizing the precise role of experience on development. With VR, researchers can systematically alter the experiences available to animals and measure which experiences change core mental skills. Since the vast majority of studies have tested subjects raised in natural worlds, we do not currently know the extent to which initial knowledge is shaped by postnatal learning. By raising animals in VR worlds, we can push perception and cognition beyond their natural bounds and explore the role that experience plays in the development of knowledge.

VR-based controlled rearing can also be used to explore the robustness of a range of core mental skills because many properties of natural environments can be manipulated in VR (*67*). Researchers have used VR to alter object physics (*68*), the scale of one's body size in space (*69*), depth cues (*70*), the nature of navigable environments (*71, 72*), and optic flow (*73, 74*). Future studies can leverage these techniques to probe the robustness of core abilities, like intuitive physics, navigation, numerical cognition, and social knowledge. This approach will allow researchers to characterize the nature of the core inductive biases that allow newborns to learn so rapidly and efficiently about the world.

**Implications for biologically inspired artificial intelligence.** The brain is often conceptualized as a "continual learning system" that learns throughout an organism's lifespan (*75*). While the brain may be able to learn some abilities at any age, this view obscures the possibility that some conceptual structures are robust to change, either due to sensitive periods for learning, or constraints on the learning mechanisms (e.g., STDP). Our study suggests that some initial knowledge, once acquired, is robust to change against opposing experience, thereby scaffolding and constraining visual learning during early postnatal development.

Building AI systems that learn like humans and animals may thus require initial 'prenatal' training phases, where artificial neural networks are grown and trained on internally generated prenatal data (e.g., retinal waves) using biologically plausible learning rules (e.g., STDP) and critical periods. This prenatal training would create a robust computational core for guiding 'postnatal' learning about the world, potentially improving the learning efficiency of AI systems. Accomplishing this goal of building a newborn-like computational core will require future work charting which knowledge exists at birth and whether that initial knowledge is flexible or rigid.

In sum, we demonstrated that object permanence is present and robust during the earliest stages of postnatal development. Our results inform the classic nature/nurture debate, showing that newborn brains have a robust object-like inductive bias at the onset of vision. This bias allows animals to solve object permanence tasks without extensive experience with objects. This bias also impedes animals from learning about unnatural events that violate object permanence. Our results support the general view that newborn animals enter the world with inductive biases that shape how even their earliest sensory experiences are processed by the brain. It remains to be seen how general this ability is across animals that mature more slowly and learn more flexibly.

—————————————

**Methods and Materials**
*Subjects*

Sixteen Rhode Island Red chicks of unknown sex were tested (Exp. 1: n = 8; Exp. 2: n = 8). No subjects were excluded from the analyses. The eggs were obtained from a local distributor and incubated in darkness in an OVA-Easy incubator (Brinsea Products Inc., Titusville, FL). To avoid exposing the chicks to uncontrolled visual input, we used night vision goggles to transport the chicks in darkness from the incubation room to the animal chambers. Each chick was raised in its own chamber. This research was approved by University of Southern California's Institutional Animal Care and Use Committee.

*Controlled-Rearing Chambers*

The chicks were raised in automated controlled-rearing chambers (66 cm length × 42 cm width × 69 cm height) constructed from white, high-density plastic. The chambers contained no real-world (solid, bounded) objects. To present object stimuli to the chicks, virtual objects were projected on two display walls situated on opposite sides of the chamber. The display walls were 19" liquid crystal display monitors (1440x900 pixel resolution). Food and water were provided *ad libitum* in transparent troughs in the ground. We used grain as food because a heap of grain does not behave like an object (i.e., a heap of grain does not maintain a rigid, bounded shape). The floors were constructed from wire mesh supported by transparent beams. Micro-cameras in the ceilings, and custom-designed tracking software, were used to track the chicks' behavior.

*Experiment 1 Procedure*

In the Input Phase (Days 1-4), we presented newborn chicks with a virtual imprinting stimulus: a red 3-lobed object that moved around a gray textured, virtual room (SI Movie 1). Other than the imprinted object, there were no objects in the controlled-rearing chamber or in the virtual room, so the chicks never saw an object occluded by another object. The object stimuli were presented as a sequence of 1-min videos. During each video, the object either 1) moved to different locations in the chamber (translation videos) or 2) moved to the left or right side of the chamber and then rotated in place (rotation videos). On the translation videos, the object repeatedly moved to positions that were short, medium, or far distances away from its current position. On the rotation videos, the object moved to the left or right side of the virtual room (the same positions used for the screens in the Test Phase), performed 4½ rotations, and then moved back to the center of the chamber.

In both the translation and rotation videos, the object started and ended in the same position and orientation, allowing the 1-min videos to be concatenated together to produce a video of a continuously moving object. The translation and rotation videos were presented in 10-min blocks. The virtual object switched monitors every 2 hours. When the object was presented on one monitor, the other monitor contained an identical virtual room (but with no object).

In the Test Phase (Days 5-10), we presented two types of test trials. On the "object movement" trials (SI Movie 2), the test video started with the imprinted object in the middle of the room. Two screens (blues cubes with diagonal dark blue stripes) then rose from the floor of the virtual room over 4 s. Next, the imprinted object turned toward one of the screens (4 s) and moved behind the screen (6 s) until the object was fully covered by the screen. When the object was fully occluded, we started the 2-min search period.

On the "invisible displacement" trials (SI Movie 3), the test video started with the imprinted object in the middle of the room, followed by the rising of the two screens (4 s). After 3 s, one of the screens then moved to the center of the room until it fully covered the imprinted object (5 s). After another 3 s, the screen (and the hidden object) moved back to the screen's original location (5 s).

To measure the chicks' object permanence, we used a chick-directed design. Specifically, we used the Unity game engine to trigger test trials to start based on the chick's position in the chamber. At the start of each trial, the imprinted object appeared on the opposite side of the chamber from the chick. When the chick moved across the chamber and entered the zone next to the object, the chick triggered the test trial. Thus, the chicks controlled the presentation of the test trials. Because the chicks initiated the trials, each chick completed a different number of trials (mean number of trials per chick = 1,351; SE = 189). The trial type (object movement vs invisible displacement) and the eventual location of the object (left vs right screen) was fully crossed and randomized across trials.

*Experiment 2 Procedure*

We used the imprinted object from Exp. 1 and a white background. In the Input Phase (Days 1-4), the chicks were reared in either the natural world (SI Movie 4) or unnatural world (SI Movie 5). In the natural world, the object always moved behind and reappeared from behind the same screen, whereas in the unnatural world, the object always moved behind one screen and reappeared from behind the opposite screen.

In both worlds, the chicks were presented with a sequence of 1-min videos in which the imprinted object repeatedly moved in and out of view from behind the two screens. Each video started with the object at the middle of the screen, rotating around a frontoparallel vertical axis at the rate of 1 rotation every 5 s. Two screens (gray rectangles) then rose from the floor on either side of the object (5 s). The screens remained stationary and the object continued to rotate for 5 s. Then, the object moved towards one of the screens until it was completely hidden behind that screen (8 s). After hiding, the screens then lowered back into the floor (5 s to lower), revealing the object. Across trials, we randomized whether the object moved behind the left vs. right screen and the amount of time that elapsed (0, 5, 10, or 20 s) before the screens dropped to reveal the imprinted object. In the Input Phase, the chicks were presented with thousands of these hiding events.

In the Test Phase (Days 5-8), we used the chick-directed procedure to present test trials. The test trials were identical to the animations presented in the Input Phase, except that the screens never dropped to reveal the location of the object (SI Movies 6-7). As in Experiment 1, we measured the chicks' behavior for 2 min. after the object moved out of view. For the test trials, we randomized whether the object hid behind the left or right screen. We also randomized whether the test screens were the same color (1/2 trials) or a different color (1/2 trials) as the screens used in the Input Phase. The average number of trials collected per chick was 1,661 (SE = 104).

**Acknowledgments:** This research was approved by the University of Southern California Institutional Animal Care and Use Committee.

**Funding:**
NSF CAREER grant BCS-1351892 (JNW). James McDonnell Foundation Understanding Human Cognition Scholar Award (JNW). Jacobs Foundation Fellowship (TDU). DARPA Machine Commonsense Reasoning grant (TDU, ES). Center for Brains, Minds and Machines (TDU, ES).

**Author contributions:**
Conceptualization: JNW, TDU, ESS
Methodology: JNW, TDU, BWW
Data Collection: JNW, BWW, SMW
Data Analysis: SMW8